\documentclass[prd,twocolumn,superscriptaddress,showpacs]{revtex4}
\usepackage{graphicx}
\usepackage{float}
\usepackage{dcolumn}
\usepackage{float}
\usepackage{bm}
\usepackage{mathrsfs}
\usepackage{ulem}
\usepackage[per-mode = fraction]{siunitx}
\usepackage{hyperref}
\usepackage{xcolor}
\usepackage{soul}
\usepackage{amssymb}
\usepackage{amsthm}
\usepackage{mathtools}

\newcommand{\YRS}{YbRh$_2$Si$_2$ }
\newcommand{\YNP}{YbNi$_4$P$_2$ }

\begin{document}

\title{Concise guide for electronic topological transitions}
\author{A.A.~Varlamov}
\affiliation{CNR-SPIN, c/o Dipartimento DICII, Universitá ``Tor Vergata'', Viale del Politecnico 1, I-00133 Rome,
Italy}
\author{Y.M.~Galperin}
\affiliation{Department of Physics, University of Oslo, 0316 Oslo, Norway}
\affiliation{A.F.~Ioffe Physico-Technical Institute of Russian Academy of Sciences,  Polytekhnicheskaya 26, 194021 St.~Petersburg, Russia}	
\author{S.G.~Sharapov}
\affiliation{Bogolyubov Institute for Theoretical Physics, National Academy of Science of
Ukraine, 14-b Metrologicheskaya Street, Kiev, 03143, Ukraine}
\author{Yuriy Yerin}
\affiliation{ Dipartimento di Fisica e Geologia, Universitá degli Studi di Perugia, Via Pascoli, 06123 Perugia, Italy}

\date{\today }

\begin{abstract}
In this short review we pass through the milestones in the studies of the electronic topological transitions (ETT) and focus on some recent applications of the ideas worked out in their classical theory. These are: two-dimensional electron systems, de Haas--van Alphen effect, classification of ETT in multidimensional systems, superconductivity in systems close to ETT, thermoelectricity in heavy-fermion systems, where the cascades of topological changes of Fermi surface (FS) are generated by magnetic field. The history of studies of ETT is inextricably linked with Kharkov school of condensed matter physics, with such names as I.M.~Lifshitz, 
V.G.~Bar'yakhtar and many other. Among them is Moisey Isaakovich Kaganov, who contributed much in studies of the role of geometry and topology of FS in physical properties of the metals. Two of the authors (A.V. and Y.G.) had a honor and pleasure to work  with ``Musik'', as all friends called 
Kaganov~\cite{Blanter1994}; all of us have been learning the niceties of science from his books.  “The Fermi surface is the stage on which the drama of the life of the electron is played out” wrote Kaganov and Lifshitz. We devote this work to their memory. 
\end{abstract}
\maketitle

\section{Origin and adolescence of ETT}

The discovery of a surprising new phase transition in liquid helium
by W.H. Keesom and coworkers in Leiden in 1932 who
observed a lambda-shaped “jump” discontinuity in  
the temperature dependence of the specific heat followed by 
the first classification of the general types of transition between phases of matter, 
introduced by Paul Ehrenfest in 1933 (see the historical overview \cite{Jaeger1998}). The proposed classification of phase transitions is based on the discontinuities of the derivatives of the free energy as a function of the parameter governing transition. The lowest derivative of the free energy that is discontinuous at the transition point labels the kind of the transition in this scheme.  First-order phase transitions exhibit a discontinuity in the first derivative of the free energy with respect to some thermodynamic variable. Second-order phase transitions are continuous in the first derivative but exhibit discontinuity in a second derivative of the free energy. Under the Ehrenfest classification scheme, there could in principle be third, fourth, and higher-order phase transitions.
This classification later was applied to the study of the transition from the normal to the superconducting state in metals,  discussing liquid-gas, order-disorder, and  paramagnetic-ferromagnetic  phase transitions. 

As is known, the shape and size of the Fermi surface (FS) may change under the influence of such factors as uniform compression, changes in the concentration of ingredients in the alloy and anisotropic deformation \cite{Varlamov1989}. In particular, changes may occur in the topology of the FS: individual voids may appear and disappear, or a closed FS may transform into an open one. In 50-ies  Mott and Jones~\cite{Mott1936} made a general observation that singularities in the electron state density can manifest themselves in certain characteristics of a metal. Later on, Van Hove~\cite{VanHove1953} noted that the quasiparticle density of states in a crystal  possesses singularities at the energies corresponding to  change in the FS topology.
The anomalies associated with the $2\frac{1}{2}$-order phase transition were discussed by Ziman~\cite{Ziman1956} when he tried to explain the behaviour of the electrical resistance of copper at low temperature. 
Finally, in the  seminal paper \cite{Lifshitz1960} I.M.~Lifshitz
consistently formulated  the idea of an electron 
topological transition in 1960, and it was 
this paper that has stimulated further experimental and theoretical studies in this direction.

I.M.~Lifshitz noticed that changes in topology manifest themselves as certain anomalies in thermodynamic  and galvanomagnetic characteristics of the metal. He performed a theoretical study of these anomalies and found that, at zero temperature and in the absence of impurity electron scattering, the topological changes occur when the Fermi energy reaches a certain critical value $\epsilon_c$. In this case, if the energy of quasi-particles,  $\epsilon(\mathbf{p})$, has an absolute extreme, a void is either formed or disappears, whereas if $\epsilon_c$ is a hyperbolic singular point, then at $ \epsilon= \epsilon_c$ there occurs a disruption of the ``neck'' that connects two parts of the FS, and an open FS may become closed.
\begin{figure}[th]
\includegraphics[width=.8\columnwidth]{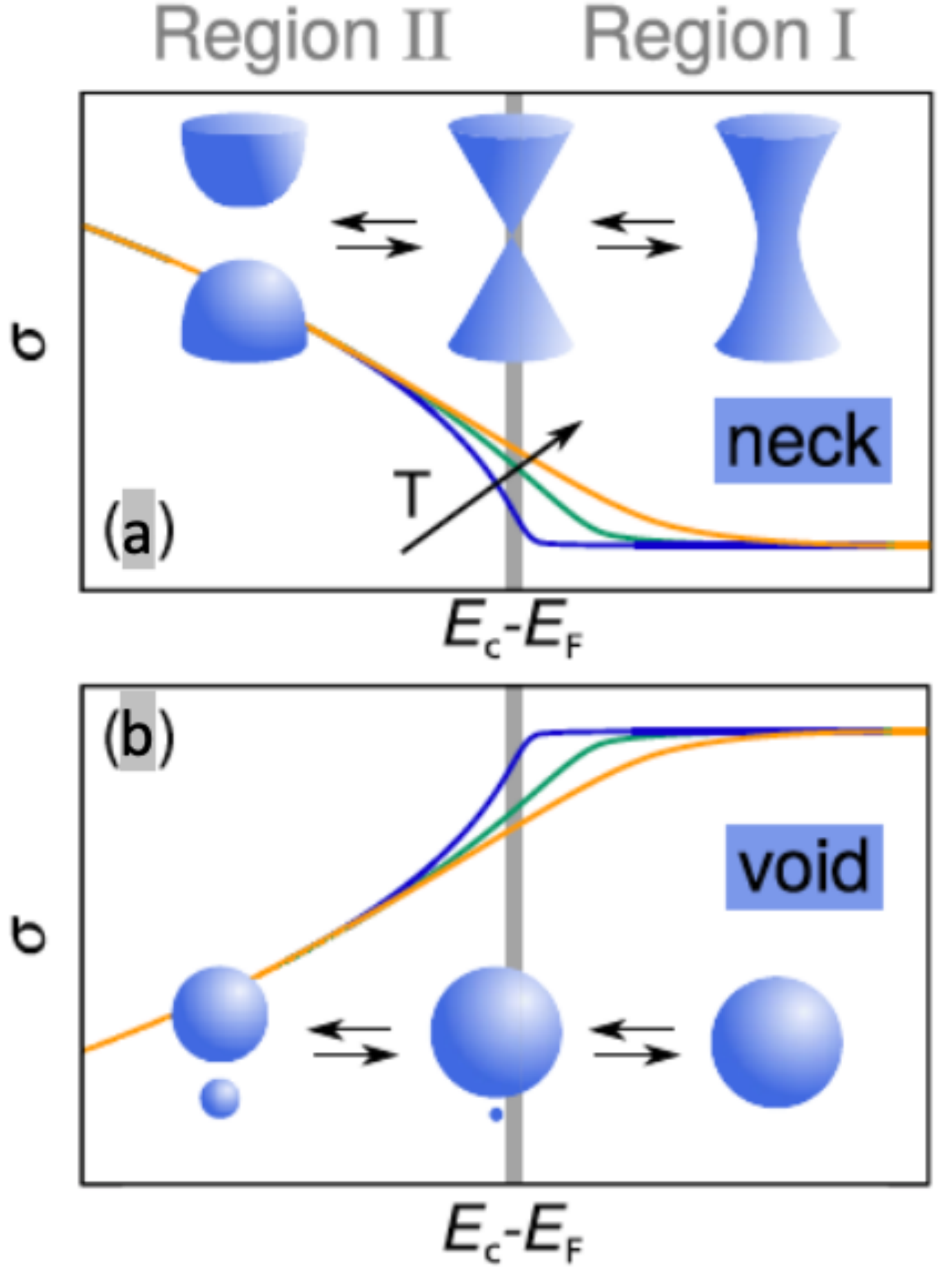}
\caption{ Panels present two type of Lifshitz transition: neck disruption and void formation.  The plots for conductivity are reproduced from \cite{Pfau2017PRL} and are based on 
Refs. \cite{Varlamov1985,Varlamov1989} for the clean case and are presented for three 
different temperatures (a more smooth curve corresponds to the higher
temperature)}.
\label{fig_1}
\end{figure}

Near such transition in 3D metal and at $T=0$ the density of electron states can be presented as a sum~\cite{Lifshitz1960}
\begin{equation}
\label{DOS_ETT}
\nu(\epsilon)=\nu_0 +\frac{(2m_1m_2m_3)^{1/2}}{\pi^2\hbar^3}|\epsilon-\epsilon_c|^{1/2}\Theta \left[\pm(\epsilon-\epsilon_c)\right]
\end{equation}
of a smooth function $\nu_0$ and an additional term $\delta \nu(\epsilon)$ associated with the transition, which is non-zero only on one side of critical point $\epsilon_c$: on those  where either a new void is formed or a neck is disrupted. Here $m_i$ are modules of the principal values of an effective mass tensor 
and $\Theta$ is the Heaviside function.

One can easily check that such discontinuity in the derivative of the 
density of states results in the appearance at $T=0$ of a non-smooth contribution to the thermodynamic potential:
\begin{equation}
\delta \Omega(z) \sim |z|^{5/2}\Theta \left(\pm z\right)
\end{equation}
where $z=\epsilon-\epsilon_c$ is the energy parameter driving transition. It characterizes the closeness of the system to the point where the number of components of topological connectivity of the FS changes. This parameter can be biased by hydrostatic pressure, anisotropic deformation, doping by some isovalent impurities, magnetic field etc. 

Evidently, the second derivatives of the thermodynamic potential, i.e. specific heat capacity, diamagnetic susceptibility, have a vertical kink $\sim |z|^{1/2}\Theta \left(\pm z\right)$. The same non-smoothness appears also in conductivity \cite{Lifshitz1960}
\begin{equation}
\delta \sigma(z) \sim |z|^{1/2}\Theta \left(\pm z\right).
\end{equation}

In their turn,  the third derivatives (for instance, the coefficient of thermal expansion) from one side of the transition point demonstrate singularity $\sim |z|^{-1/2}$.  It is from this observation I.M.~Lifshitz  has named such electron transitions at $T = 0$ and in the absence of impurity scattering as the $2 \frac12$-order phase transitions. 

At finite temperature however (or when there are impurities that scatter electrons) the singularity at $z = 0$ becomes smeared out, so, strictly speaking, in this case we cannot call such topological transformation of the FS a phase transition. To distinguish these cases from  the true $2 \frac12$-order phase transition, we will call the latter as Lifshitz, or an electronic topological transition (ETT).

Following the prediction by I.M.~Lifshitz of a possibility of changing the FS topology and related  effect on the physical properties of metals had drawn the attention of both theoreticians and experimentalists  (for detailed review see \cite{Varlamov1989,Blanter1994} and references therein). Yet, the low temperature  measurements of such tiny changes in heat capacity, or precise study of the coefficient of thermal expansion were found to be rather  complicated, and this fact decelerated the studies of electronic topological transitions.

As an important milestone in this initial period of the studies of Lifshitz transitions should be mentioned the paper by  Baryakhtar and Makarov~\cite{BM1965}. The authors studied the effect of 
FS topology change under the effect of pressure on the superconducting transition temperature and found that the latter exhibits a considerable nonlinearity. Up to the present the discovery of the $T_c(P)$ anomaly (see, e.g., \cite{BM1972,Higgins1998,Lazarev1966,Ignat1968}) is considered as an experimental proof of the Lifshitz transition existence.

It is necessary to mention also the paper by
M.A.~Krivoglaz and Tuo Hao, who had extended the Lifshitz transition thermodynamics for the case 
of metal with impurities ~\cite{Krivoglaz1966}.

\section{Discovery of the giant anomaly in Seebeck coefficient and recursion to topological transitions}
\label{sec:Seebeck}

\begin{figure}[th]
\includegraphics[width=.99\columnwidth]{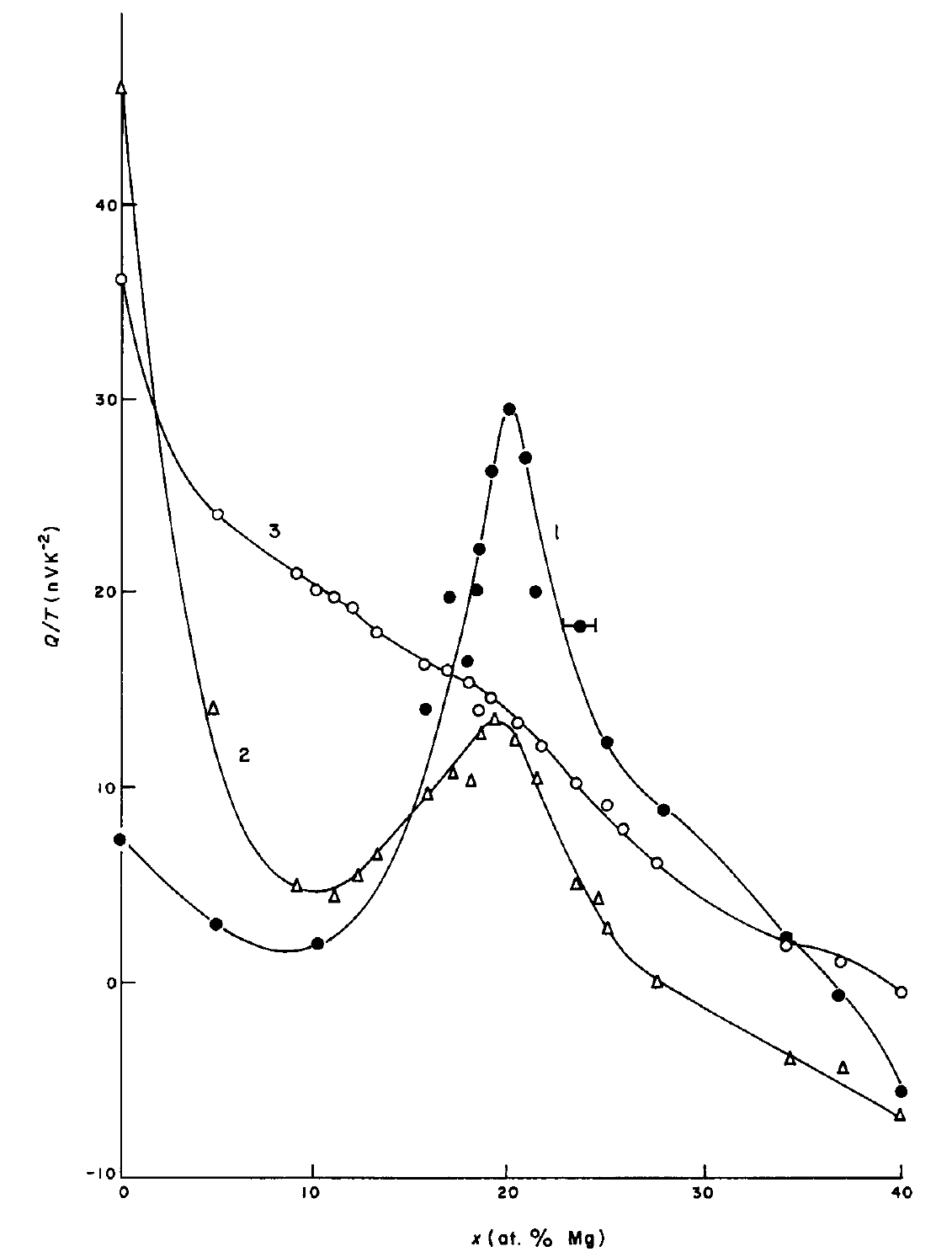}
\caption{The dependence of thermopower on magnesium concentration at different temperatures $T=4.2$ K (filled circles), $T=78$ K (triangles) and $T=4.2$ K (blank circles) \cite{Egorov1983}.}
\label{thermopower_Egorov}
\end{figure}

The next stage in the studies of the electronic topological transitions started at the beginning of 80-ies and was related to the discovery~\cite{Egorov1983} of the giant peak in Seebeck coefficient of the alloy $ \text{Li}_{1-x} \text{Mg}_x$   when the concentration of magnesium approached $x$=0.2 at. $\%$ (see Fig. \ref{thermopower_Egorov}). In according to the band calculations namely this value corresponds to the touching of the alloy's FS the Brillouin zone edge, i.e. transition from the closed  FS to the open one \cite{Vaks1981}.  It was observed as a small
kink at room temperature, which developed in the well pronounced peak at nitrogen temperatures with the further growth up to 30 times with respect to the background at the temperature of liquid helium.

It is worth to recall that this discovery was preceded by the prediction of the anomaly in thermoelectric coefficient based on the some naive explanation \cite{Vaks1981}. The authors related the kink 
feature
in the density of states ($\delta \nu \sim \sqrt{z})$  (see Eq.~(\ref{DOS_ETT})) to 
the non-smoothness in the mean free path of the charge carriers,
\begin{equation}
l(\bf{p})=l_0 (\bf{p})+|z|^{1/2}\Theta \left(\pm z\right)l_1(\bf{p}),
\end{equation}
where $l_0(\bf{p})$ is the mean free path of electron scattering at the mainland of the FS with
$\bf{p}$ being the quasiparticle momentum and $l_1(\bf{p})$  is the term related to the singular region.

Then they used the Mott's formula for the Seebeck coefficient $S=\pi^2 T/(3e)d (\ln \sigma)/d \epsilon |_{\epsilon=\epsilon_F}$ with $\sigma = \sigma_0+\delta \sigma(z)$.
Accordingly, $|z|^{-1/2}$ singularity  in thermoelectric coefficient
in this approach occurs  due to disruption of the neck.
Such additive treatment assumes that the electrons in the both pieces of the FS are independent. 
Yet, a more detailed consideration raises  questions: (i)
which value of the
electron velocity has to be used to calculate  $\delta \sigma(z)$? 
If one chooses the value of the velocity at the just disrupted neck,
then its smallness would suppress the gained singularity.
(ii) What is the value  of the relaxation time, corresponding to the electrons scattered from 
the new piece of the FS?

The answers to these questions were given in the papers~\cite{VP84,Varlamov1985,AP86,VP88}, where the scattering of the electrons belonging to the FS with variable number of the components of topological connectivity was studied. The authors demonstrated that in the case of the ETT of the type of new void formation three scenarios of the electron scattering on impurities are possible. The first one is trivial: an electron 
from the mainland returns to the mainland
after the scattering from impurity.
Corresponding probability in the case of isotropic scattering is determined by the Fermi golden rule:
\begin{equation}
\tau_0^{-1}= 2\pi n_{\mathrm{imp}} \nu_0 |U|^2,
\label{golden}
\end{equation}
where $n_{\mathrm{imp}}$ is the impurities concentration and $U$ is the amplitude of the scattering potential, which we assume independent of momentum. More precise analysis shows, that this probability depends on the electron energy $\omega = \epsilon - \epsilon_F$, but very weakly: the correction to Eq. (\ref{golden}) is $ \sim \tau_0^{-1}\omega/\epsilon_F$. Namely this smallness is responsible for the standard weakness of the Seebeck effect ($S \sim T/\epsilon_F$) in normal metals. 

The second possibility for electrons after the scattering from impurity
is to travel from the main part of the FS to 
the small void.
Probability of such process is definitely suppressed by the smallness of the density of states in the new void (i.e., available states for newcomer in result of scattering process). However, this smallness for the Seebeck coefficient is over and above compensated by the strong energy dependence of this contribution. The last 
scenario of impurity scattering for  electron from 
the new void is to return back. This process can be neglected  due to  
smallness of the corresponding density of states which 
enters quadratically in the corresponding probability.

As a result, the total relaxation time can be written as 
\begin{equation}
\tau_{(3)}^{-1}(\omega,z)\!=\! \frac{\tau_0^{-1}}{2 \sqrt {\epsilon_F}}\left[ \kappa \left( \pm \omega - \epsilon_F\right) \mp \kappa \left( \pm \omega \pm z \right),  \right]
\label{taugen}
\end{equation}
where the function
\begin{equation}
\kappa(x)= \sqrt{2} \left[\left( \frac1{4 \tau_0^2} +x^2 \right)^{1/2}-x \right]^{1/2}
\label{Krivoglaz}
\end{equation}
appeared firstly in the paper of Krivoglaz  and Tu Hao~\cite{Krivoglaz1966}. 
The choice of signs in Eq.~(\ref{taugen}) is the following: sign ``+'' corresponds to the transition of the neck disruption type (hyperbolic point) while sign ``-'' corresponds to the formation of the new void.
More explicit form Eq.~(\ref{taugen}) acquires in the clean limit, when $ \tau_0^{-1} \ll T$:
\begin{equation}
\tau_{(3)}^{-1}(\omega,z)\!=\! \tau_0^{-1}\left[1 \!-\! \frac{\omega}{\epsilon_F}\!-\!\left(\frac{|z+\omega|}{\epsilon_F} \right)^{1/2} \Theta(\!-\!z\!-\!\omega)\right].
\end{equation}
 Equation~(\ref{taugen}) allows to present the Seebeck coefficient for the system close to ETT in the compact form valid for the wide range of impurity concentrations ($ \tau_0^{-1} \ll\epsilon_F$) and temperatures:
\begin{equation}
\label{single-peak-Seebeck}
S(T,z) \sim \frac{k_B }{eT^2 \tau_0}\int_{-\infty}^{\infty}  \frac{ \omega d\omega}{\cosh^2\frac{\omega}{2T}} \tau(\omega,z).
\end{equation}
 \begin{figure}[bh]
\includegraphics[width=.8 \columnwidth]{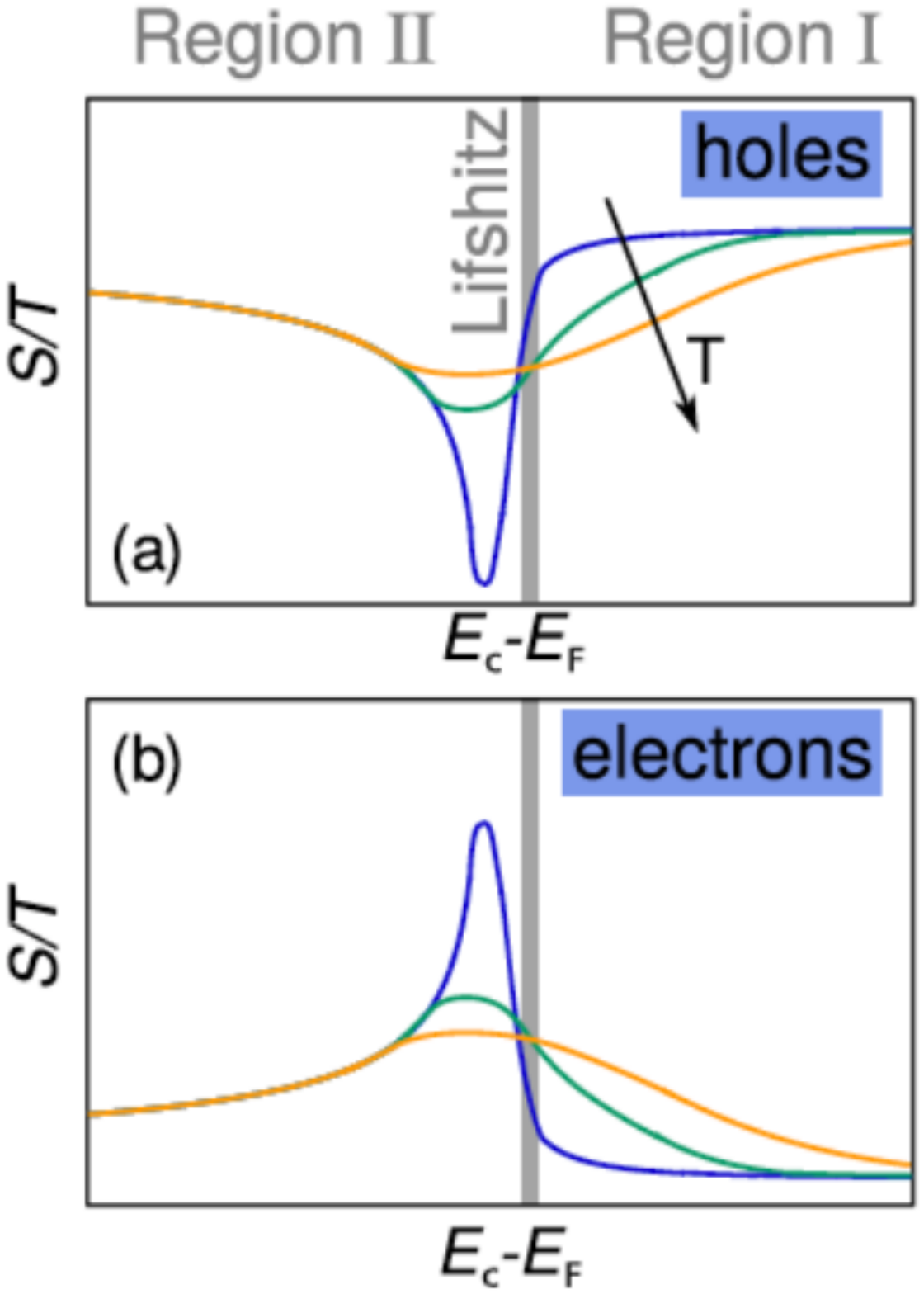}
\caption{Panels present theoretical calculations of the signatures 
of the Lifshitz transition in thermopower $S$. The plots are reproduced from \cite{Pfau2017PRL} and are based on  Refs. \cite{Varlamov1985,Varlamov1989} for the clean case and presented for three different temperatures. The sign of the thermopower (maximum or minimum) is defined by the type of charge carrier
(see Fig. \ref{fig_1}).}
\label{Fig_2}
\end{figure}
It is essential that consideration of such problem in the model of neck disruption reproduces the same result for Seebeck coefficient with replacement $z \rightarrow -z$ and $\omega \rightarrow -\omega$, the latter means the replacement of electrons by holes. Therefore the shape of singularity close to ETT depends only on the character of topological changes of the FS but not on its geometrical type, viz. void or neck.

The discovery of singularity in Seebeck coefficient in $\text{Li}_{1-x}\text{Mg}_x$ alloy (see Fig. \ref{thermopower_Egorov}) became a trigger for intensive study of transport phenomena in vicinity of very different topological transitions, generated by isotropic pressure, anisotropic deformation, inducing of the isovalent impurities and other methods (see detailed references in~\cite{Varlamov1989,Blanter1994}).

\section{ETT in two dimensions}

\subsection{Dissimilarity with  the three-dimensional case}

End of the XX century was marked by the special interest in low dimensional systems, this is why let us briefly review  now  the  2D  case (we will follow Ref.~\cite{BTV94}).  It turns out that the density of states singularities, oppositely to the 3D case, are different here depending on the transition type.
In the case of a void formation and an elliptical surface
of the spectrum, the non-smooth
part of the density of states (DOS) turns out to be proportional to $\Theta \left( { \pm \left( {\varepsilon  - {\varepsilon _c}} \right)} \right)$, while in the case of neck disruption the DOS singularity is proportional to $\ln \left( {\varepsilon  - {\varepsilon _c}} \right)$~\cite{N1966,MA1979,SK1970}. 
As a consequence a singular part of the thermodynamic
potential is proportional either
to ${z^2}\Theta \left( { \pm z} \right)$ (void formation) or ${z^2}\ln \left| z \right|$ (neck disruption) at zero temperature and in absence of a random potential (the
role of the impurity scattering was discussed in Ref.~\cite{AKR1991}.

One can see that these two expressions are very different from each other. In the case of neck disruption the singularity is nonzero from both sides of the transition and it is even on $z$. This fact is a direct consequence of the above mentioned symmetry of a system corresponding to simultaneous exchange of holes and electrons branches in the spectrum and a change of $z$ sign (see the discussion below Eq. (\ref{single-peak-Seebeck}).
In this case the second derivative of the
thermodynamic potential has a logarithmic singularity at the transition point.

In the case of a void formation the DOS becomes nonzero from one side of the transition only (those corresponding to the new void). First derivatives of the thermodynamic potential have kinks at $z=0$, and
the second ones break.

In accordance with Ehrenfest terminology, both two-dimensional void formation and neck disruption must be referred as second order phase transitions. However, it is important to note that this terminology
seems to be too general and can give rise to some misunderstanding.
In fact, in the current literature, the term ``second order phase transition" is strictly associated with the phenomena described by the Landau theory, where it was supposed that the dependence of the thermodynamic potential on the order parameter is different, above and below $T_c$, due to the special choice of the temperature dependence of  the coefficient in the quadratic term. So, in spite of the common features between the types of transitions, the 2D void formation and neck disruption have to be differentiated from the second-order phase transitions in
the Landau sense.
\begin{figure}[th]
\includegraphics[width=0.99 \columnwidth]{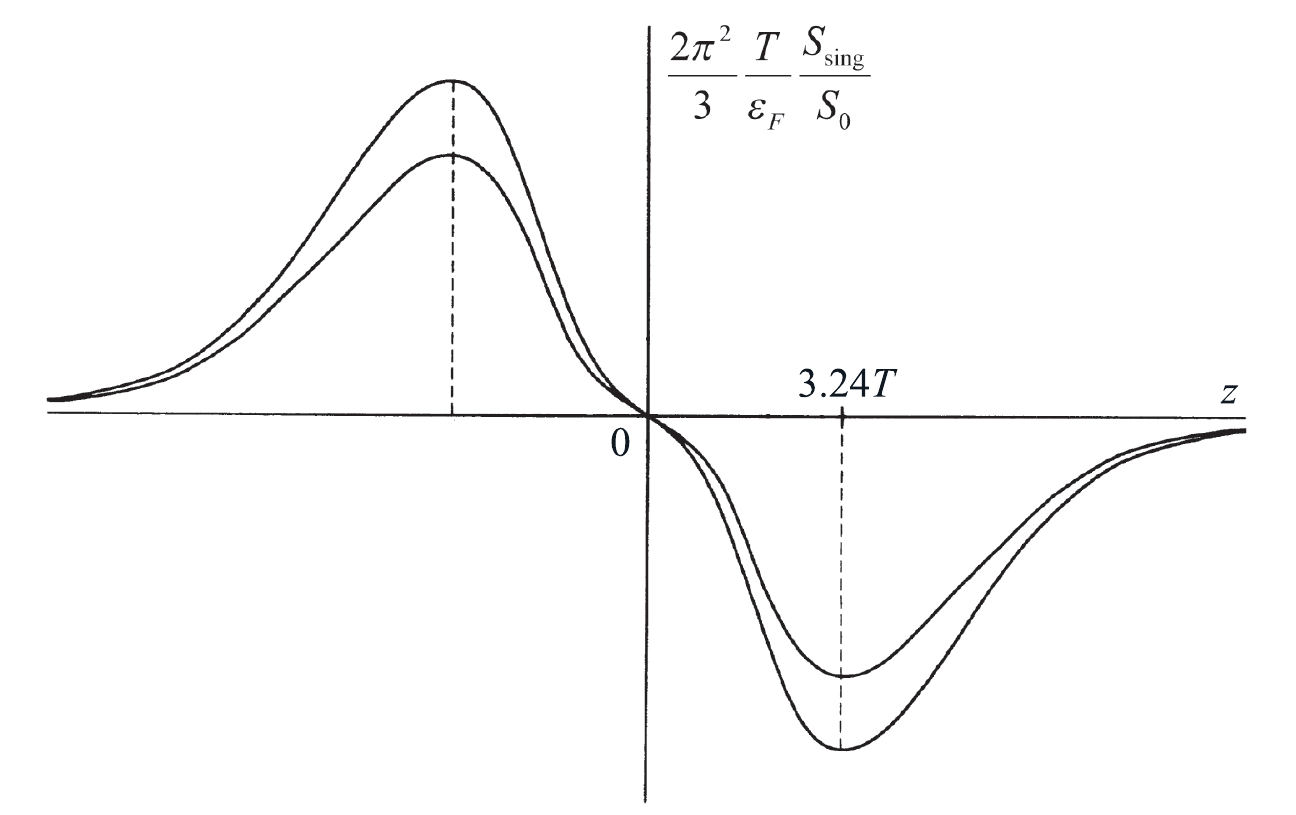}
\caption{Theoretical calculations of the thermal power $S(z)$ for
the case of the Lifshitz transition of the neck disruption type in two dimensions.
The plot is reproduced from Ref. \cite{BPV92}, two curves correspond to different values of electron concentrations.}
\label{2dTHEP}
\end{figure}

\subsection{Thermopower close to the 2D ETT of the neck disruption type}
	
Slight differences between the two aforementioned ETT lead naturally to some differences for the thermodynamic and kinetic properties. The two most remarkable examples are the thermal expansion coefficient
and thermoelectric power. The first one does not exhibit any singularities at all at the point of void formation and exhibits a sharp peak in the temperature dependence (with height proportional $\mu/T$) in
the case of neck disruption \cite{N1966}. Thermopower dependence
versus $z$ exhibits a sharp asymmetric peak with height $ \sim \mu /T$ in the case of void formation (in the 3D case it was $(\mu/T)^{1/2}$) )\cite{AKR1991,Z2}.

Special attention in 2D systems requires the case of neck disruption.  Here, unlike the 3D case, the system acquires a new type of symmetry: the relaxation time remains unchanged when one replaces the electrons by holes and changes the sign of $z$:
\begin{equation}
\label{2Drelaxtime}
\tau_{(2)}^{-1}(\omega,z)=\frac2{\pi}\tau_0^{-1}\ln \frac{4 \epsilon_F}{|\omega+z|}.
\end{equation}
Correspondingly, this fact dramatically affects the shape of singularity in Seebeck coefficient: instead of well pronounced asymmetry with which we are already familiar for both types of transition in 3D case and void formation in 2D case,  in the case of 2D neck disruption the dependence $S(z)$ is even, with a minimum and maximum of the same height \cite{Z2,BPV92} (see Fig. \ref{2dTHEP}).

\subsection{Topogical transitions in a quasi-2DEG:
	Quantization of entropy}\label{sec_qent}

The characteristic example of the ETT in two dimensions is the filling by electrons of the empty subband formed due to the size quantization in the quasi-two-dimensional  electron gas (quasi-2DEG).   Each intersection of the levels of electron size
quantization by the chemical potential (when $\mu$ passes the level $E_N$), can be considered as the point of
the Lifshitz phase transition of the void formation type, where the FS acquires a new component of topological connectivity. 

In gated structures, a series of Lifshitz transitions can be controlled by the gate voltage.
Corresponding anomalies in the thermodynamic and transport characteristics, in particular, thermoelectric
coefficient related to the peculiarities of the energy dependence of the
electron momentum relaxation time have been studied experimentally and
theoretically in Refs.~\cite{Z1,Z2} and \cite{BPV92,AKR1991}, respectively.  In particular,
ETTs spectacularly  manifest themselves in behavior of \textit{differential entropy (entropy per electron)}, $s\equiv V^{-1}(\partial \mathcal{S}/\partial n)_T$ ( $\mathcal{S}$ is the entropy, $V$ is the volume of the system which below we will assume to be unit).  
An elegant way to measure this quantity directly was recently demonstrated~\cite{Pudalov}.

As was shown~\cite{Varlamov2016,Galperin2018}, the quantization of the energy spectrum of quasi-2DEG into subbands leads to a very specific quantization of the entropy: $s$ exhibits sharp maxima as the
chemical potential $\mu $ passes through the bottoms of size quantization
subbands ($E_{i}$). 
\begin{figure}[th]
	\includegraphics[width=0.7\columnwidth]{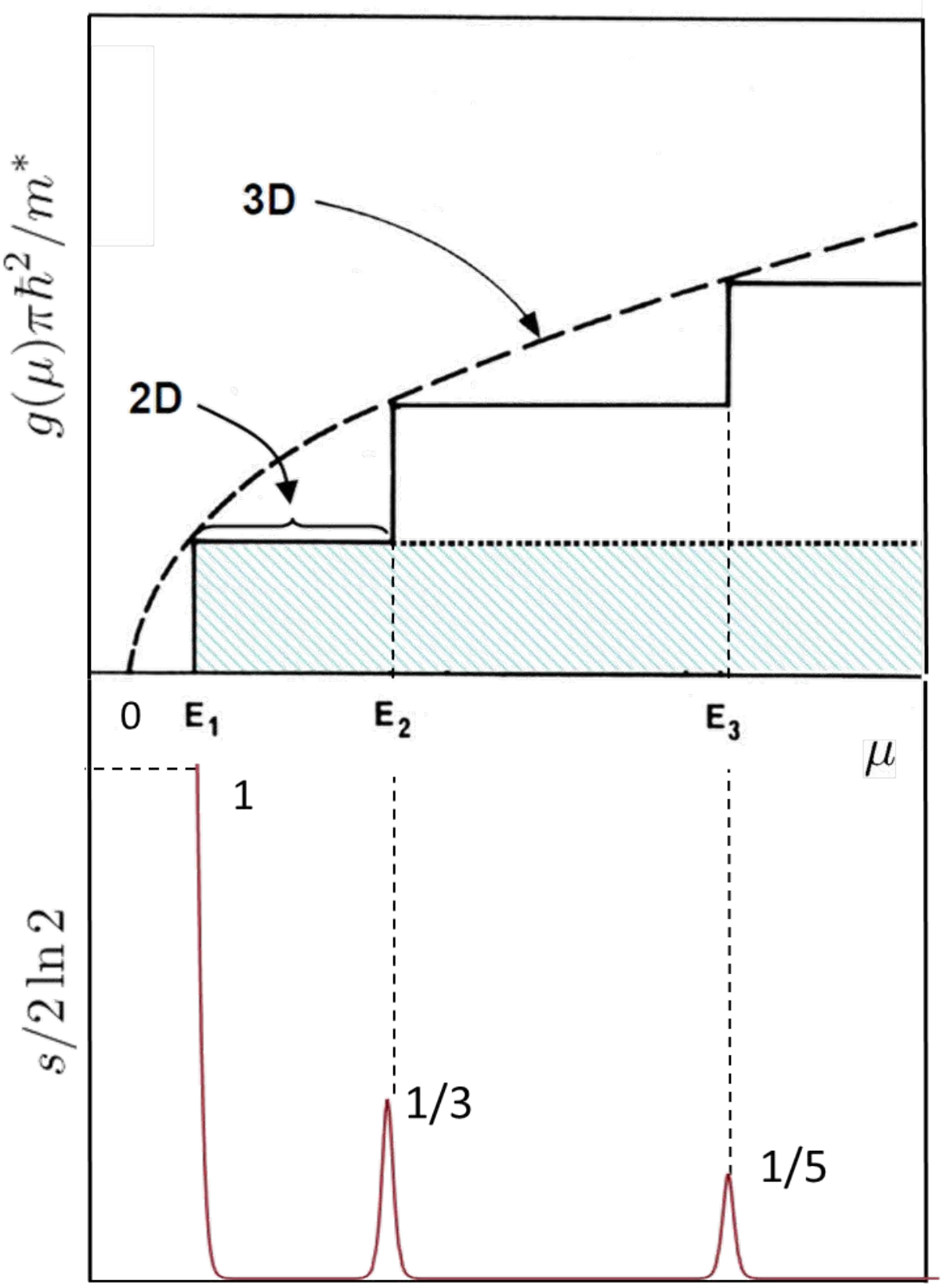}
	\caption{Schematic representation of the dependencies of the electronic
		density of states (upper panel) and the partial entropy (lower panel) as
		functions of the chemical potential. The plot is reproduced from \cite{Varlamov2016}.}
	\label{fig1}
\end{figure}
Indeed, the density of electronic states (DOS) in a
non-interacting 2DEG and in absence of scattering has a staircase-like shape~\cite{Ando}: 
\begin{equation}
\nu(\mu )=\frac{m^{\ast }}{\pi \hbar ^{2}}\sum\limits_{i=1}^{\infty }\Theta
\left( \mu -E_{i}\right) ,  \label{nuclean}
\end{equation}%
with $m^{\ast }$ being the electron effective mass.
The differential entropy, $s$, can be found using
the Maxwell relation 
\begin{equation}
s=\left( \frac{\partial \mathcal{S}}{\partial n}\right) _{T}=-\left( \frac{\partial
	\mu }{\partial T}\right) _{n}=\left( \frac{\partial n}{\partial T}\right)
_{\mu }\left( \frac{\partial n}{\partial \mu }\right) _{T}^{-1}.
\label{fullderiv}
\end{equation}%
The relationship between the electron concentration $n$, chemical potential $%
\mu $ and temperature $T$ can be found integrating Eq.~(\ref{nuclean}) over
energies with the Fermi-Dirac distribution: 
\begin{equation}
n\left( \mu ,T\right) =\frac{m^{\ast }}{\pi \hbar ^{2}}\sum\limits_{i=1}^{%
	\infty }\int\limits_{0}^{+\infty }\frac{{\Theta}(E-E_{i})}{\exp \left( 
	\frac{E-\mu }{T}\right) +1}\,dE.  \label{ngen}
\end{equation}%
Using Eq.~(\ref{ngen})
one can express 
$s$ in the form of sums
over the subband levels averaged over energy with temperature
smearing factor. 

In the absence of scattering this result is independent of the shape of the
transverse potential that confines 2DEG and of the material parameters
including the electron effective mass and dielectric constant. The value of the entropy in the $N$-th maximum depends 
\textit{only} on the number of the maximum, $N$: 
\begin{equation}  \label{s}
s\vert_{\mu = E_n} \equiv \left(\frac{\partial \mathcal{S}}{\partial n}\right)_{T,\,
	\mu = E_n} = \frac{\ln 2}{N - 1/2}\, .
\end{equation}

The universality of the above quantization rule can be broken both by
disorder and by electron-electron interactions. Using a simple model of
Lorentzian smearing of the electronic spectrum one can see that it leads to the
relative correction of $\sim 1 /T\tau $, where $\tau $ is the electron
life time. For the case of a single-band 2DEG the role of the
electron-electron interaction in the partial entropy was investigated in~%
\cite{Pudalov}.

The variation of differential entropy $s$ versus chemical potential is
schematically shown in Fig.~\ref{fig1}, lower panel. The peaks in it correspond to the steps of the density of states shown in the upper panel of the same figure. 
This dependence can be
interpreted in the following way. At low temperatures, the main contribution
to the entropy is provided by the electrons having energies in the vicinity
of the Fermi level, the width of the `active' layer being $\sim T$. If the
electron DOS is constant within the layer then by adding an electron one
does not change the entropy. Hence, the entropy is independent of the
chemical potential, $(\partial S/\partial n)_{T}\rightarrow 0$. However, if
the bottom of one of the subbands falls into the active layer, the number of
`active' states becomes strongly dependent on the chemical potential. In
this case, adding an electron to the system, one changes the number of
`active' states in the vicinity of the FS. Consequently, the
partial entropy strongly increases. The peaks of the partial entropy
correspond to the resonances of the chemical potential and electron size
quantization levels. The further increase of the chemical potential brings
the system to the region of the constant density of states, where the
partial entropy vanishes again.

At $T\rightarrow 0$ (yet $T\gtrsim \hbar /\tau $ ) the peaks of $s$ are
located at $\mu \rightarrow E_{N}$, $N>1$, the maximal values being $s_{\max
}(N)=\ln 2/(N-1/2)$. At finite $T$ the peaks acquire finite widths of the
order of $T$ and shift toward negative values of $\mu -E_{N}$ analogously to the shift of the maximum in the Seebeck coefficient (see Fig. \ref{Fig_2}).

\section{ De Haas-van Alphen effect as the 3/2 order phase transition}

\begin{figure}[th]
\includegraphics[width=.8\columnwidth]{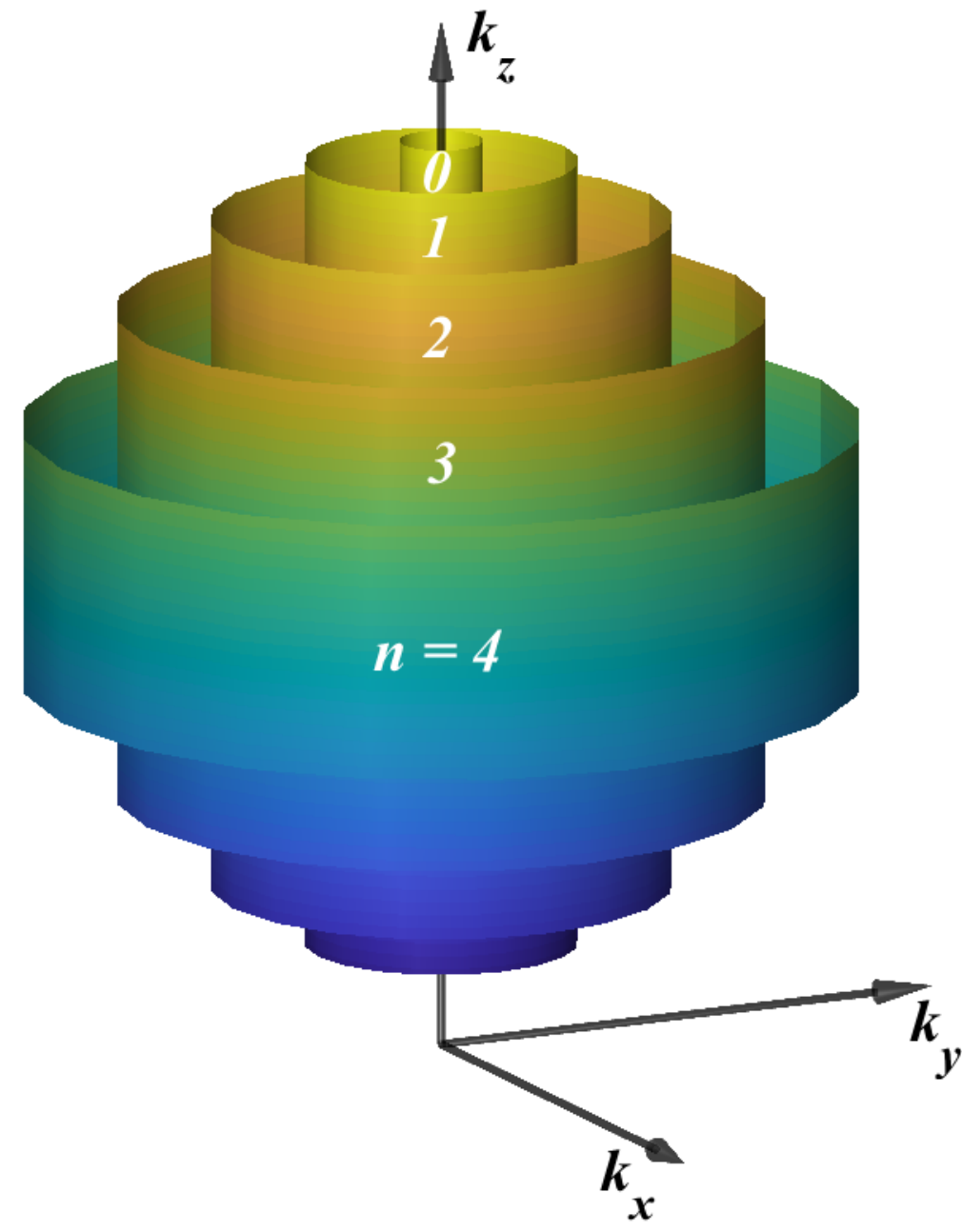}
\caption{Fermi sphere in magnetic field is transformed in a set of Landau cylinders, corresponding to the filling by electrons of the degenerated states at the Landau levels.}
\label{FS_LL}
\end{figure}

Let us discuss the situation when the magnetic field $H$ is applied to the metal and the cross-section of the FS in the direction perpendicular to $H$ is closed (we will follow Ref.~\cite{BTV94}). The motion of the electrons is quantized in this plane while remains free and effectively one-dimensional
in the magnetic field direction, viz. because he energy spectrum depends on one momentum projection only.
It is important to note that the intersection
of the sequent Landau level and chemical potential
level (see Fig.~\ref{FS_LL}) can be treated, in certain sense, as a $1\frac12$ order ETT.
Indeed, if one denotes Landau level energies (which can be found from the  Lifshitz-Onsager quantization rule for reasonable fields) as 
${E^{\left( n \right)}}$, then the DOS singularity in the vicinity of 
this crossing reads~\cite{LAK73}
\begin{equation}
\label{DOS_variation}
\delta \nu\left( \varepsilon  \right) \sim {\left( {\varepsilon  - {E_n}} \right)^{ - 1/2}}\Theta \left( {\varepsilon  - {E_n}} \right).
\end{equation}
It is clear from the last expression that the number of available
for electrons states for $n$-th level is non-smooth and proportional to
${\left( {\varepsilon  - {E_n}} \right)^{1/2}}\Theta \left( {\varepsilon  - {E_n}} \right)$ and the non-smooth part in the thermodynamic
potential turns out to be
\begin{equation}
\delta \Omega \left( \varepsilon  \right) \sim {\left( {\mu  - {E_n}} \right)^{3/2}}\Theta \left( {\mu  - {E_n}} \right)
\end{equation}
at zero temperature.

We have to note that the chemical potential depends on the magnetic field due to the particle number conservation law; nevertheless this dependence
becomes important in the ultra-quantum limit only, which is 
beyond the scope of our discussion here \cite{LAK73,AS64,A69}.

It can be concluded that the de Haas-van Alphen oscillations arising when the next Landau level passes through the chemical potential value, can be treated as ETT to which the 1½ order has to be attributed. The second derivatives of the thermodynamic potential tend to infinity at $T=0$ if $z(H)=E_n$, and at finite temperatures exhibit sharp peaks with magnitude ${\left( {\mu /T} \right)^{1/2}}$.
This singularities are nothing else but inverse magnetic field oscillations of magnetic susceptibility in the De Haas-van Alphen effect. It is noteworthy that the kinetic coefficients behave exactly as  ETT 
in 3D case: conductivity (Shubnikov-de Haas effect), 
thermopower \cite{VP89,PV89}, etc.

\section{Multi-dimensional ETT}

After this review of the different phenomena showed by the electronic topological transitions, we would like to mention some general features characteristic for such transitions and to classify them formally
for arbitrary space dimensionalities $n$ (we will follow Ref.~\cite{BTV94}).

In the neighborhood of any non-degenerate critical point, the spectrum can be described by a quadratic form and the equation of the Fermi hypersurface can be expressed as
\begin{equation}
\label{hypersurface}
\sum\limits_{i = 1}^n {\frac{{p_i^2}}{{2{m_i}}}}  = \epsilon  - {\epsilon _c}.
\end{equation}
The shape of this hypersurface is determined by the signature of $\left\{ {{m_i}} \right\} = 1,2,...,n$ and by the sign of $\epsilon  - {\epsilon _c}$. The topological type of transition (only the sign of
$\epsilon  - {\epsilon _c}$ is changed), is established by the signature of $\left\{ {{m_i}} \right\}$ only.

Let us start the discussion from the 0D case. Here the Fermi sea of electrons may be formally treated as a point which can be empty ($\epsilon < \epsilon_c$) or occupied by two electrons ($\epsilon > \epsilon_c$). Therefore, only the transition which can be formally treated as the void formation type, can occur in this case (the void does not exist or occupies all 0D space).

In the 1D case, the Fermi sea is presented by the interval $\left[ { - {p_F},{p_F}} \right]$. These two limiting points play the role of the Fermi "surface". For the surface
\begin{equation}
\frac{{p_x^2}}{{2{m_x}}} = \epsilon  - {\epsilon _c}
\end{equation}
only two signatures, $[ + ]$ and $[ - ]$, are possible and both correspond to void formation (of electron or hole types). So in this case, like in the 0D case, only one type of transition, namely void formation, is possible. The same void formation type transition is typical for the electron system of any dimensionality and
corresponds to the signatures $\left\{ { +  +  + ... + } \right\}$ and
$\left\{ { -  -  - ... - } \right\}$. But already in the 2D another type of signature: $\left\{ { +  - } \right\}$ (or $\left\{ { - +} \right\}$) for the quadratic form Eq. (\ref{hypersurface}) appears and establishes side by side with the void formation, another type of ETT - the neck disruption. A similar type of transition occurs in the 3D space for the signatures $\left\{ { +  +  -} \right\}$ and $\left\{ { -  -  + } \right\}$; the same as for higher space dimensions.

In the 4D space, besides the two transitions of the mentioned above nature, the new transition type corresponding to the symmetrical signature $\left\{ { +  +  - - } \right\}$ appears.

In the general case of a $n$-dimensional space $n + 1$ different signatures are possible of which $\left\{ { +  +  + ... + } \right\}$ and $\left\{ { -  -  - ... - } \right\}$ describe a new electron and hole void formation, respectively, $\left\{ { +  +  + ... + -} \right\}$ and $\left\{ { -  -  - ... - +} \right\}$ correspond to electron/hole neck disruption and the other are associated with more complex topological transformations. In general, the surface containing $p$ pluses and $q$ minuses ($p, q > 1$) can be imagined as $(p + q)$ dimensional hyperboloid
and $\frac{1}{2}n + 1$ (if $n$ is even) or $\frac{1}{2}\left( {n + 1} \right)$ (if $n$ is odd) different kinds of transitions are possible.

It can be seen that if $n$ is even ($n=2k$) one among these transition types (which corresponds to the symmetrical signature of $k$ positive and $k$ negative masses) possesses a remarkable symmetry: the system the system does not vary if all electrons are interchanged with holes and the sign $\epsilon  - {\epsilon _c}$ changes also. This symmetry (it was already described for the 2D system) must imply the DOS singularity dependence to be even function of $\epsilon  - {\epsilon _c}$. All other types of transitions do not exhibit such a symmetry and we cannot
expect consequently that the singular part of the DOS will be even function of energy. It is worthwhile to stress that such electron-hole symmetry exists for even dimensional spaces only. For odd dimensions,
the two hyperbolic surfaces beyond and above the transition are not topologically equivalent.

The critical exponent of the DOS 
 non-smoothness may be easily calculated for the $n$-dimensional case of void formation. Really, the DOS is proportional to (see~\cite{LAK73})
\begin{equation}
\nu^{[n]} \left( \epsilon  \right) \sim \int \! {\delta \left[ {\varepsilon  - \varepsilon ( {\mathbf{p}} )} \right]} {d^n}{\mathbf{p}}
=\frac{d}{{d\epsilon }}\int \! {\Theta \left[ {\epsilon  - \epsilon ( {\mathbf{p}} )} \right]} {d^n}{\mathbf{p}},
\end{equation}
where the index $[n]$ denotes the dimensionality of the space. 
One can see that the last integral  presents nothing
else than a volume inside the equal energy surface ${\epsilon \left( {\mathbf{p}} \right)}=\epsilon$ and is proportional to the number of electron states. The singular part of it is proportional to the volume of the $n$-dimensional ellipsoid,
\begin{equation}
N^{[n]}(\epsilon) = \frac{\prod\limits_{i = 1}^n {{p_i}}}{2^n \pi^{n/2}\hbar^n \Gamma(n/2+1)},
\end{equation}
where $p_i$ are the large semi-axis of the ellipsoid in momentum space. Due to the homogeneity of the quadric form in Eq. (\ref{hypersurface}), the
volume of the $n$-dimensional ellipsoid is proportional to ${\left( {\epsilon  - {\epsilon _c}} \right)^{n/2}}\Theta \left( {\epsilon  - {\epsilon _c}} \right)$. So the non-smooth part of the DOS 
is proportional to $dN^{[n]}/d\epsilon$, i.e.,
\begin{equation}
\label{}
\delta \nu^{[n]}(\epsilon)\sim {\left( {\epsilon  - \epsilon_c} \right)^{n/2 - 1}}\Theta \left( \epsilon  - \epsilon_c \right).
\end{equation}
The contribution to the
thermodynamic potential reads
\begin{equation}
\delta \Omega^{[n]}\left( \epsilon  \right) \sim {\left( {\epsilon  - \epsilon_c} \right)^{n/2 + 1}}\Theta \left( \epsilon  - \epsilon_c \right).
\end{equation}
 We can conclude now that $n$-dimensional formation of a void can be considered as phase transition of $n/2+1$ order at zero temperature, and for high dimensions ETT-related singularities will be manifested weakly. This conclusion concerns other types of transitions too.

\section{Revisited ETT in superconductors}

It has been shown that abrupt topological modification of the FS can be achieved by tuning of different macroscopic parameters such as temperature, pressure, strain, external magnetic fields and doping.
\begin{figure}[th]
\includegraphics[width=.92\columnwidth]{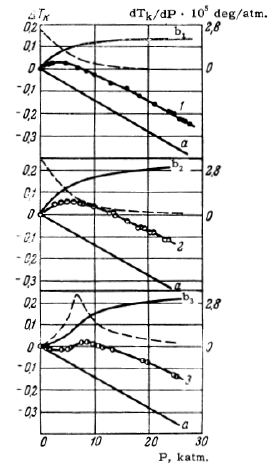}
\caption {Dependence of the change of superconducting transition temperature in thallium and its alloys as a function of the pressure in kilo-atmospheres
for pure Tl (up, curve 1), Tl with 0.45 $\%$  at.(atomic percentage)
Hg (middle, curve 2) and Tl with 0.9 $\%$  at. Hg (bottom, curve 3). Curves $a$ represent linear contribution, while $b_1$, $b_2$ and $b_3$ are nonlinear components of curves 1, 2, 3, respectively. The dashed lines are the nonlinear parts of the dependence of $dT_c/dP$ on the pressure. The plot was reproduced from Ref.~\cite{BM1965}.}
\label{delta_Tc_exp}
\end{figure}
A dramatic reshaping of the FS topology viz. Lifshitz transition often leads to the emergence of the plenty phenomena also in the superconducting phase of a material and as a result to the unusual behavior and the alteration of many properties relevant to superconductivity. 

On of the first evidence of the Lifshitz transition in a superconductor was revealed in the above mentioned paper of Baryakhtar  and  Makarov \cite{BM1965} where  the effect of pressure on the superconducting transition temperature $T_c$ in various metals was studied both theoretically and experimentally. It was found that with the increasing of the pressure the critical temperature of pure thallium and of thallium with Hg impurities  undergoes a nonlinear variation (see Fig.~\ref{delta_Tc_exp}.).

Interpretation of such anomalous behaviour can be done within the BCS theory of  superconductivity. 
The corresponding equation for the temperature 
dependent energy gap $\Delta (T)$ can be written in the following form:
\begin{equation}
\label{BCS_gap_deform}
1 = \lambda \! \int\limits_{ - {\omega _D}}^{{\omega _D}}\frac{\nu ( \xi )\,  d\xi}
{2\sqrt {\xi ^2 + {\Delta^2( T )} }}
\tanh \left( {\frac{{\sqrt {{\xi ^2} + {\Delta ^2}\left( T \right)} }}{2T}} \right),
\end{equation}
where $\lambda$ is the constant of electron-phonon interaction, $\omega_D$ is the Debye frequency, and $\nu(\xi)$ is the density of states. 
In the vicinity of ETT $\nu(\xi)$ is given by Eq.~(\ref{DOS_ETT}).

Theoretical analysis of the Eq.~(\ref{BCS_gap_deform}) for limiting cases  $T=0$ and $T=T_c$ allows to interpret 
qualitatively
the experimental data shown in Fig.~\ref{delta_Tc_exp}.  At the
relatively low pressures the Fermi energy $\epsilon_F$ is slightly higher than $\epsilon_c$. Until the difference $\epsilon_F-\epsilon_c$  remains less than $\omega_D$  the critical temperature increases due to the presence in Eq. (\ref{BCS_gap_deform}) of the correction $\delta \nu (\xi)$ which strongly varies as a function of energy. With the further increase of pressure, corresponding growth is saturated, yet the other mechanisms enter in interplay: the non-monotonic behaviour of the critical temperature is connected with the dependence on pressure of other parameters of $\omega_D$, $\nu_0$, and $\lambda$ and the presence of impurities (see Fig.  \ref{delta_Tc_exp}, curves 2 and 3) \cite{Lazarev1966, Ignat1968}.

Later the same theory was applied successfully for interpretation of the behavior of the critical temperature as a function of pressure in In-Cd~\cite{Higgins1998}, in In-Sn and In-Pb alloys~\cite{Zavaritskii1971, Smith1973}.

The similar qualitative dependence of $T_c$ as a function of lattice strain was detected in strontium ruthenate $\text{Sr}_\text{2}\text{RuO}_\text{4}$ that probably exhibits an exotic, odd-parity kind of superconductivity \cite{Steppke2017}. Compressing the $a$ axis of the $\text{Sr}_\text{2}\text{RuO}_\text{4}$ lattice drives the critical temperature through a pronounced maximum, that is a factor of 2.3 higher than $T_c$ of the unstrained material.  The combination of experimental data and theoretical calculations gives evidence that the observed maximum $T_c$ occurs at or near a Lifshitz transition when the Fermi level passes through a Van Hove singularity \cite{Steppke2017,Sunko2019, Barber2019}.

It is worth note that after discovery of the copper-based high-temperature superconductivity there was an attempt to exploit the theory of ETT for the elucidation of the order parameter symmetry in such compounds \cite{Abrikosov1994}. The model proposed by A. Abrikosov was based on two pillars: i) existence of a large dielectric constant associates with the ion cores that leads to a weak screening of the Coulomb forces and corresponding long range interactions of conduction electrons; ii) existence of the extended saddle point singularities and their dominant contribution to formation of the order parameter. Assumption that the density of states and energy gap $\Delta (p)$ in the singular region of the momentum space to be much larger than beyond, results in the modification of the BCS self-consistency Eq. (\ref{BCS_gap_deform}). Its solution demonstrates the possibility of the  $s$-wave and $d$-wave order parameter symmetries coexistence, the absence of the Josephson effect in the $c$-oriented junctions based on such compounds, their tunneling conductance as well as the enhancement of the critical temperature.

Despite the fact that this model is rather simple and succeeded to explain most of the apparently contradictory properties of high-temperature superconductors at the moment of its proposition, it has not become widely accepted.

In 2008 high-temperature superconductivity
was surprisingly discovered in iron-based materials. The presence of iron, for long time believed as an enemy of superconductivity, led to the explosive growth in the study of basic physical properties of this new family and comprehension of origin of the unconventional pairing mechanism in these high temperature superconductors.  It was demonstrated by ARPES and other techniques that a common feature of iron-based superconductors is the multiple-band electronic structure. Some of these bands are very shallow with Fermi energies of several meV~\cite{Liu2010, Liu2011, Miao2014, Okazaki2014}. This circumstance allows to deplete such bands by means of doping or pressure, and therefore to induce the Lifshitz transition in such materials. As a result, complex multi-band topology gives rise to a much richer nomenclature of effects in the vicinity of ETT in these superconductors in comparison with their conventional counterparts.

One of the such an unusual phenomenon was predicted in a regime of BCS-BEC (Bose-Einstein) crossover in a two-band superconductor when the chemical potential $\mu$ is tuned in a narrow energy range around the edge $E_{\text{edge}}$ of the second band \cite{Innocenti2010}.

\begin{figure}[th]
\includegraphics[width=.6\columnwidth]{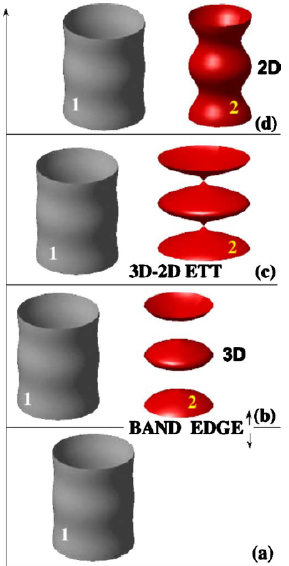}
\caption {Sketch of the evolution of the FS for a two-band superconductor  by shifting the chemical potential from  (a) to (d) so that it crosses two Lifshitz ETT. The first ETT occurs moving the Fermi level across the band edge $E_{edge}$ of the second band so that the superconductivity goes from the single FS in (a) with a single condensate to the two FS in (b) with two condensates: the first (1) has a 2D topology and the second (2) has a 3D topology. Changing $E_F$ the system crosses the critical energy $E_{3D-2D}$ where the second FS undergoes a 3D-2D ETT shown in
panel (b) changing its topology: the second closed 3D FS (b) becomes the tubular 2D FS in (d). The first large 2D FS (1) remains nearly constant when the chemical potential is shifted. Taken from Ref. \onlinecite{Innocenti2010}.}
\label{FS_ETT_2band}
\end{figure}

Shifting the chemical potential for the superconductor one can track the evolution of FS topology and the emergence of several Lifshitz transitions (Fig.~\ref{FS_ETT_2band}).  When $\mu$ is increased above $E_{\text{edge}}$, and the system undergoes an ETT with the opening of a new 3D FS (see Fig. \ref{FS_ETT_2band} a,b). When $\mu$ reaches a higher energy threshold, the electronic structure undertakes a second ETT, the 3D-2D ETT, where one FS changes topology from 3D to 2D, i.e., from “spherical” to “cylindrical” or vice versa (see Fig.~\ref{FS_ETT_2band} c,d). This ETT is a feature of stacks of metallic layers, multilayers, or so-called superlattices of quantum wells, and therefore it is typical for all existing high-temperature superconductors and novel materials synthesized by material design in the search for room temperature superconductivity.

In the general case of multi-band metal with deep primary bands and a shallow one, when the bottom of the shallow band crosses the Fermi level, ETT causes anomalous density of states with the qualitatively different behaviors on the two sides of the Lifshitz transition in the superconducting phase \cite{Koshelev2014}. The mean-field theory gives the expression for the resulting density of states
\begin{eqnarray}
{\nu _s}\left( E \right) &=& \frac{{{{\left( {2m} \right)}^{3/2}}}}{{8{\pi ^2}}}\operatorname{Re} \left[ {\sum\limits_{\delta  =  \pm 1} {\left( {\frac{{\left| E \right|}}{{\sqrt {{E^2} - \Delta _0^2} }} + \delta \operatorname{sgn} \left( E \right)} \right)} } \right. \nonumber \\ &&
\left. { \times \sqrt {\mu  + \delta \sqrt {{E^2} - \Delta _0^2} } } \right],
\label{DOS_multiband}
\end{eqnarray}
where $\Delta_0$ is the pairing amplitude (gap function) induced in the shallow band by Cooper pair exchange with the deep bands. Eq. (\ref{DOS_multiband}) was obtained within conditions that energy gaps in each band are much smaller than the Fermi energies for the deep bands, while the relation between $\Delta_0$ and $\mu$ can be arbitrary. Also, the term corresponding to the contribution to pairing from the shallow band was excluded in self-consistent equations for energy gaps.

It is important to note that the term with $\delta=-1$ contributes to Eq.~(\ref{DOS_multiband}) only if $\mu>0$ and $\left| E \right| < \sqrt {{\mu ^2} + \Delta _0^2} $.  At $\Delta _0=0$ Eq.~(\ref{DOS_multiband}) yields the normal density of states,
\begin{equation}
\label{DOS_multiband_normal}
{\nu _n}\left( E \right) = \frac{{{{\left( {2m} \right)}^{3/2}}}}{{4{\pi ^2}}}\sqrt {E + \mu } \ \Theta \left( {E + \mu } \right),
\end{equation}
while in the limit $\mu  \gg {\Delta _0}$ the density of states recovers the standard symmetric BCS shape
\begin{equation}
\label{DOS_classic}
{\nu _s}\left( E \right) = {\nu _n}\left( 0 \right)\frac{{\left| E \right|}}{{\sqrt {{E^2} - \Delta _0^2} }}.
\end{equation}

Equation (\ref{DOS_multiband}) describes main peculiar features of a multiband superconductor near the Lifshitz transition. On one side of the ETT the density of states diverges at the energy equal to the induced gap, whereas on the other side it vanishes.

\section{Cascade ETT in heavy fermion systems}	

The specific feature of the extensively studied since 2006 heavy-fermion systems
is that the Lifshitz transitions are  driven by 
a magnetic field (see, e.g.,  Ref.~\cite{Pourret2019JPSJ} for a brief overview).
In these compounds, flat quasiparticle bands with the widths often comparable to the Zeeman splitting of the energy bands $(g_{\mathrm{eff} }/2)\mu_BH$ for field of order 
\SI{10}{T} cross the Fermi level.  
For example, a series of strong anomalies in the thermoelectric power 
were recently observed both in the ferromagnetic Kondo
lattice material \YNP when the magnetic field varies in the range from 
\SI{0.4}{T} to \SI{18}{T} \cite{Pfau2017PRL} 
and in the heavy fermion compound \YRS when the field varies in the range from 
\SI{9.5}{T} to \SI{13}{T} \cite{Pourret2019JPSJ}. 

The experimental findings of Ref.~\cite{Pfau2017PRL} were theoretically
interpreted in terms  of the series of the independent Lifshitz topological 
transitions as considered above in Sec.~\ref{sec:Seebeck}. 
The sign of the thermopower signature (maximum or minimum) allows to determine 
the type of charge carrier (electrons or holes).
Moreover, resolving the shape of the peaks
it is possible to identify the type of Lifshitz transitions 
as neck or void type (see Fig.~\ref{fig_1}).

The theoretical analysis  of Ref.~\cite{Pfau2017PRL}  is valid for standing apart peaks, 
viz. when the width of the peak is much smaller than the distance between them. 
Yet, comparing the experimental situations Ref.~\cite{Pfau2017PRL}
and that one of Ref.~\onlinecite{Pourret2019JPSJ} for \YRS 
one can see that the minimal distance between the anomalies in the former 
is \SI{0.75}{T} (for the most of the peaks it is a few Tesla), while all  anomalies in \cite{Pourret2019JPSJ} experiments are separated by the value of \SI{0.6}{T}.

When the  thermopower peaks are getting close to each other one should
also include the superposition phenomena between the new pockets 
occurring in the narrow energy range. An indication that the 
the superposition of the different Lifshitz transitions indeed 
occurs in the experiments on \YRS, is that it is hard to identify
the type of the transition. 
The asymmetry of the individual peaks which represents 
the characteristic feature of  the Lifshitz transition,
is blurred in \YRS as compared with \YNP.

To take into account the specificity of a cascade of Lifshitz transitions on quasiparticle scattering the following model was suggested in \cite{Pourret2019JPSJ}.
In the review we will not go into specific detail of \YRS compound
which was also modelled by {\it ab initio} computations and 
following Ref.~\cite{Pourret2019JPSJ} represent the general idea
how to describe the cascade of Lifshitz transitions.

\begin{figure}
\includegraphics[width=0.99\columnwidth]{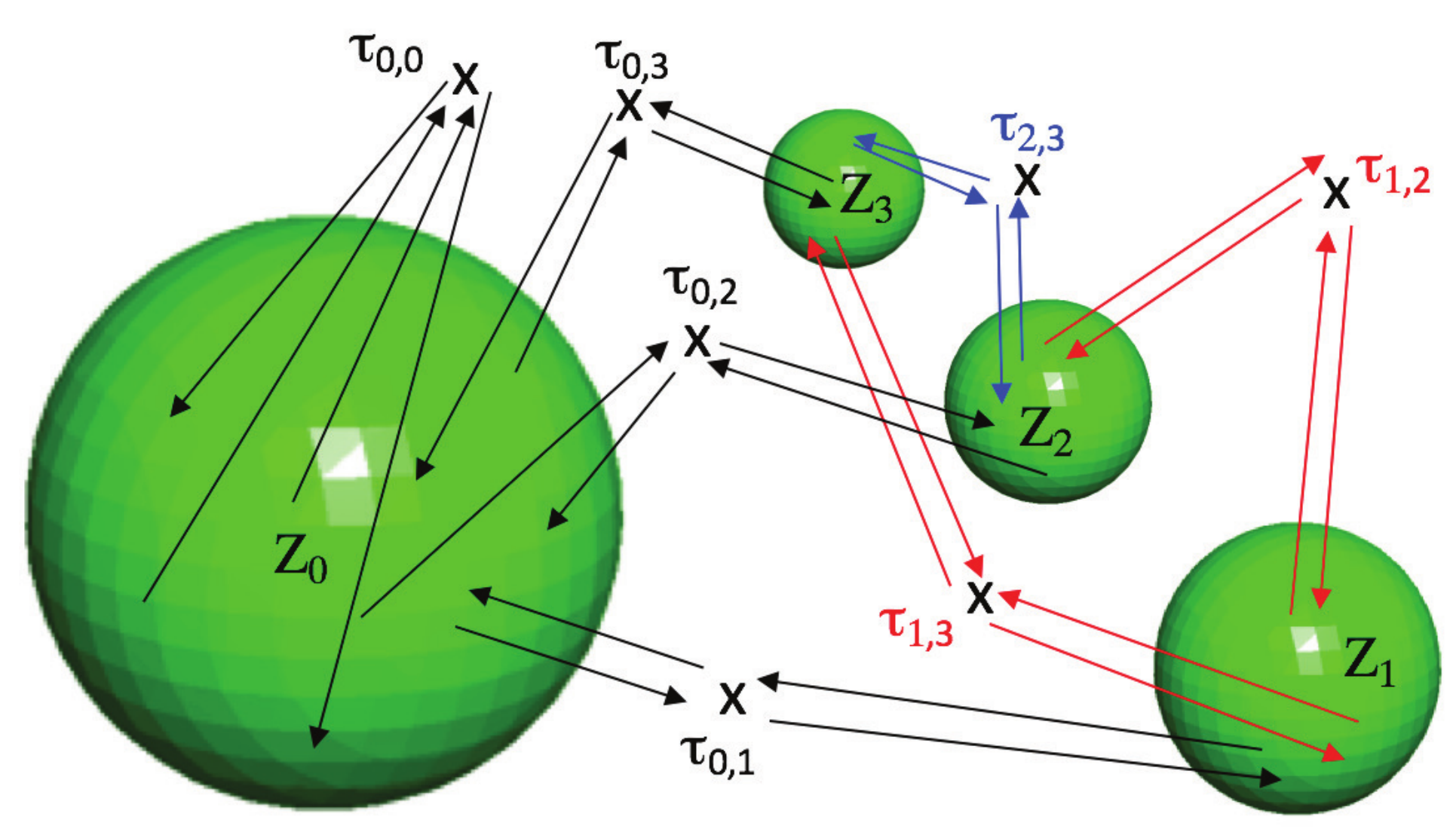}
\caption{(Color online) Schematic representation of the various quasiparticle scattering processes between the pockets with the energy sizes $Z_{i}$. $\tau_{i,j}$ correspond to the processes from pocket $i$ to pocket $j$ ($\tau_{0,j}$ refers to the process from the main FS $Z_{0}$)  through impurities (crosses). Taken from \cite{Pourret2019JPSJ}.}
\label{grocery}
\end{figure}

We  assume that all topological changes of the FS are due to the formation of new spherical pockets (see Fig.~\ref{grocery}). Here the index $i=0$ denotes to regular part of the FS hereafter referred to as mainland while the critical parts are labelled by $i=1,...N$. The quasiparticle excitation spectrum for each of them can be presented as:
\begin{equation}
\xi _{i}(\mathbf{p})=\frac{(\mathbf{p}-\mathbf{p}_{ci})^{2}}{2m_{i}}%
-Z_{i},\qquad i=0,\ldots N,  \label{spectra}
\end{equation}
where $\mathbf{p}_{ci}$ is the position of its center in the Brillouin zone, $Z_{i}=\mu -\epsilon _{ci}$ is its energetic size, $\epsilon_{ci} 
= \mu $ is the critical point of multi-valued function of energy when the i-th voids appears, $m_{i}$ is the corresponding effective mass.

In Refs.~\cite{Varlamov1985,VP88} it was demonstrated that the anomaly in the  Seebeck coefficient in the vicinity of Lifshitz transition is directly related to the specific quasiparticle scattering starting from the bulk component of the mainland FS (the energy size is $Z_{0}$) and ending at the small pocket of the FS  (characterized by the energy size $Z_{i}$). It was shown that the latter is a trap for quasiparticles,  their velocity here is small and they die away. The next scattering with the dominating probability returns the quasiparticle back to the mainland FS. Such forth and back scattering give rise to the singularity of the Seebeck coefficient. At the first glance generalization of the described scattering mechanism to the case of multiple Lifshitz transitions seems to be trivial: it is just necessary to account for the round trips of the quasiparticles from the mainland FS to all other pockets ($Z_{1}, Z_{2}, Z_{3} \ldots $), see $\tau_{0,j}$ processes in Fig.~\ref{grocery}. Yet, as will be shown below, this is not enough. An important role is also played by the ``traveling'' of the quasiparticles between the newborn FS elements, $\tau_{1,j}$ and $\tau_{2,j}$ processes in Fig.~\ref{grocery}.
The expression for the scattering time for the quasiparticles belonging to the ``$l$''-th pocket accounting for the return to the same pocket and round trips to the smaller ones with $i>l$, 
in full analogy with the results of Refs.~\cite{Varlamov1985,VP88}, is given by:
\begin{equation}
\tau^{-1} _{l}\left( \omega ,Z_{l},..Z_{N}\right) =\sum_{i=l}^{N}{\tau }
_{l,i}^{-1}(\omega ,Z_{l}),
 \label{taul}
\end{equation}
where
\begin{equation}
\tau_{l,i}^{-1}(\omega ,Z_{l})=\frac{\kappa _{l}\left(
	-Z_{i}-\omega \right) }{2{\tau}_{l0}\sqrt{Z_{l}}}.
\label{taui}
\end{equation}
Here the function
\begin{equation}
\kappa _{l}(\varepsilon )=\left[{\left(\varepsilon ^{2}+ \frac1{4 \tau_{l0}^2}\right)^{1/2}-\varepsilon}\right]^{1/2},
 \label{kl}
\end{equation}
with $\tau_{l0}^{-1}=\pi ^{-1}n_{\mathrm{imp} }|U|^{2}m^{3/2}(2Z_{l})^{1/2}$ as the relaxation time of the electron initially belonging to the l-th void and calculated in the golden rule approximation (compare with Eqs. (\ref{golden}),(\ref{Krivoglaz}). For the sake of simplicity the effective masses for electrons in all voids are assumed to be equal, $m_{i}=m$.

Finally, the expression for the Seebeck coefficient 
(\ref{single-peak-Seebeck}) becomes now
the sum over involved cascade scattering processes (marked in Fig.~\ref{grocery} by the black, red and blue arrows, respectively) with the relaxation time $\tau(\omega,z)$
replaced by the corresponding for each term time $\tau _{l}\left( \omega ,Z_{l},..Z_{N}\right)$.

\section{Conclusion}
As we have shown the field of fermiology, related to study of geometrical and topological properties of FS, founded by I.M. Lifshitz still remains hot topic and time to time emerges in different problems of condensed matter: in new families of superconductors, heavy fermion systems, low dimensional systems, etc. Considerable merit in it belongs to M.I. Kaganov, both due to his original contributions and his review article \cite{Blanter1994} which became the handbook for new generations of researchers.

\begin{acknowledgments}

A.V. is grateful to V.S. Egorov for valuable discussion and clarification of the historical circumstances. S.G.Sh. acknowledges a partial support by the National Academy of Sciences of Ukraine grant  ``Functional Properties of Materials Prospective for Nanotechnologies'' (project No. 0120U100858). Y.Y.  acknowledges  support  by  the  CarESS project. 
\end{acknowledgments}

\end{document}